\documentclass[aps, prl, amsmath, floats, floatfix, twocolumn,superscriptaddress, nofootinbib, showpacs,reprint]{revtex4-1}

\usepackage{graphicx}
\usepackage{amsmath,amssymb}
\usepackage{amsfonts}
\usepackage{xspace} 
\usepackage[usenames]{color}
\usepackage{dcolumn}
\usepackage{bm}
\usepackage{mathrsfs}

\usepackage{ulem}
\definecolor {darkgreen}{rgb}{0.2,0.7,0.2}

\begin{document}

\title{A no-go theorem for slowly rotating black holes in Ho\v rava--Lifshitz gravity}

\author{Enrico Barausse} 
\affiliation{Department of Physics, University of Guelph, Guelph, Ontario, N1G 2W1, Canada}
\author{Thomas P. Sotiriou} 
\affiliation{SISSA, Via Bonomea 265, 34136, Trieste, 
Italy {\rm and} INFN, Sezione di Trieste, Italy}
\begin{abstract}
We consider slowly rotating, stationary, axisymmetric black holes in the infrared 
limit of Ho\v rava--Lifshitz gravity. We show that such solutions do not exist, 
provided that they are regular everywhere apart from the central singularity. 
This has profound implications for the viability of the theory, considering 
the astrophysical evidence for the existence of black holes with non-zero spin.
\end{abstract}

\date{\today \hspace{0.2truecm}}

\maketitle

Black-hole (BH) spacetimes have event horizons that act as causal boundaries. 
This property is intimately related to the causal 
structure of relativistic gravity theories, such as general relativity (GR). 
It is, therefore, reasonable to ask whether BHs can actually exist 
in theories that exhibit violations of Lorentz symmetry.

Interest in Lorentz-violating (LV) gravity theories stems 
from the fact that constraints on Lorentz violations
in gravity are significantly weaker than in the matter sector, 
since the gravity sector is weakly coupled. 
In fact, Einstein-aether theory (\ae-theory), which is GR coupled to a unit timelike vector field, 
was proposed as a benchmark for quantifying Lorentz violations 
in gravity \cite{Jacobson:2000xp}. More recently, a more direct motivation 
for LV gravity came from Ref.~\cite{arXiv:0901.3775}, where a  framework  for constructing a power-counting renormalizable gravity theory was presented.
 The desirable ultraviolet (UV) behavior is achieved
by including higher-order spatial derivatives in the gravitational action, 
but only second-order time derivatives in order to avoid loss of unitarity. 
Terms with at least 2$d$ spatial derivatives, 
where $d$ is the number of spatial dimensions, 
are required in order to achieve power-counting 
renormalizability \cite{arXiv:0901.3775,arXiv:0902.0590,arXiv:0912.4757}.

The corresponding theory is known as 
Ho\v rava--Lifshitz (HL) gravity 
(see {\em e.g.~}\cite{Sotiriou:2010wn} for a brief review). The action is 
straightforwardly constructed once a preferred foliation  is imposed, 
and is invariant under the reduced set of diffeomorphisms that leave 
this foliation intact, $T\to \tilde{T}(T)$ and $x^{i}\to \tilde{x}^{i}(T,x^i)$.
In  $3+1$ dimensions one then has \cite{arXiv:0909.3525}
\begin{equation}
\label{SBPSHfull}
S_{HL}= \frac{M_{\rm Pl}}{2}\int dT d^3x \, N\sqrt{h}\left(L_2+\frac{1}{M_\star^2}L_4+\frac{1}{M_\star^4}L_6\right)\,,
\end{equation}
where $M_{\rm Pl}$ is the Planck scale, $h$ is the determinant of the metric $h_{ij}$ induced 
on the spacelike hypersurfaces and
\begin{equation}\label{HLaction}
L_2=K_{ij}K^{ij} - \lambda K^2 
+ \xi \,{}^{(3)}\!R + \eta \,a_ia^i
\,,
\end{equation}
where $K$ is the trace of the extrinsic curvature $K_{ij}$, ${}^{(3)}\!R$ is 
the Ricci scalar of $h_{ij}$, $N$ is the lapse function, $a_i=\partial_i \ln N$,
and $\lambda$, $\xi$ and $\eta$ are dimensionless parameters.  
$L_4$ and $L_6$ denote respectively collections of all the 4th-order and 6th-order operators, while
$M_\star$ is the scale that suppresses these operators. 
$L_4$ and $L_6$ contain a very large number ($\sim 10^2$) 
of operators and independent coupling parameters and this can be 
considered as an unappealing feature of the theory. 
However, 
here we will focus on the infrared (IR) limit of the theory, 
which depends only on $L_2$. 
The only feature associated with $L_4$ and $L_6$ that will concern us
is the fact that dispersion relations cease to be linear at energies around $M_\star$. 
We will also not consider versions of the theory where the action is required to satisfy 
extra restrictions and symmetries, 
see {\em e.g.}~Refs.~\cite{arXiv:0901.3775,Sotiriou:2009gy,Sotiriou:2009bx,Horava:2010zj,Vernieri:2011aa} 
and Ref.~\cite{Sotiriou:2010wn} for a brief review.

The deviations of $L_2$ from GR are measured by $|1-\lambda|$, $|1-\xi|$ and $\eta$. The limit to GR is 
not smooth, as HL gravity presents an extra scalar mode due to the reduced symmetry with respect to GR. 
This scalar mode can exhibit instabilities, have negative energy or 
get strongly coupled at unacceptably 
low energies \cite{Charmousis:2009tc,Blas:2009yd,Papazoglou:2009fj,arXiv:1003.5666}. 
However, in a significant part of the parameter space the theory is 
free from pathologies and viable in the IR \cite{arXiv:1007.3503}. 
Avoiding strong coupling imposes an upper bound on $M_\star$ \cite{Papazoglou:2009fj,arXiv:0912.0550} 
(see Refs.~\cite{Papazoglou:2009fj, Liberati:2012jf} for the implications of this bound).

If one chooses to restore diffeomorphism invariance, the scalar mode manifests as 
a foliation-defining scalar field and the IR limit of HL gravity 
can take the form of \ae-theory with the extra condition 
that the aether be hypersurface orthogonal at the level of the action \cite{Jacobson:2010mx} 
(the ``covariantization" can be extended beyond the IR limit \cite{Sotiriou:2011dr}). 
The corresponding action is
\begin{equation}
\label{actionae}
S_{\rm \ae}=\frac{M_{\rm \ae}}{2} \int d^4x \sqrt{-g}\left(-R-M^{\alpha\beta}_{\phantom{ab}\mu\nu} \nabla_\alpha u^\mu \nabla_\beta u^\nu \right),
\end{equation}
where $g$ is the determinant of the metric $g_{\mu\nu}$, 
$\nabla_\mu$ is the associated covariant derivative, 
$R$ is the Ricci scalar of this metric,
\begin{equation}
M^{\alpha\beta}_{\phantom{ab}\mu\nu}\equiv c_1 g^{\alpha\beta}g_{\mu\nu}+c_2 \delta^\alpha_\mu \delta^\beta_\nu+c_3 \delta^\alpha_\nu \delta^\beta_\mu+c_4 u^\alpha u^\beta g_{\mu\nu}\,,
\end{equation}
$c_1$ to $c_4$ are dimensionless parameters,
and locally the aether is given in terms of the foliation-defining scalar $T$ as
\begin{equation}
\label{hypers}
u_\mu=\frac{\partial_\mu T}{\sqrt{g^{\alpha\beta}\partial_\alpha T \partial_\beta T}}\,.
\end{equation}
The correspondence of parameters with action (\ref{SBPSHfull}) is
\begin{equation}
\frac{M_{\rm \ae}}{M_{\rm pl}}=\xi=\frac{1}{1-c_{13}},\quad \lambda=\frac{1+c_2}{1-c_{13}},\quad \eta=\frac{c_{14}}{1-c_{13}}\,,
\end{equation}
where $c_{ij}=c_i+c_j$. $c_4$ (or $c_1$ or $c_3$) may be set to zero 
without loss of generality as long as eq.~(\ref{hypers}) 
holds. However, we have not done so in the equations above, in order to have 
a direct comparison with general \ae-theory, in which eq.~(\ref{hypers}) 
does not hold, but instead the aether is a full-fledged vector that satisfies 
the constraint $u^\mu u_\mu=1$.

In Ref.~\cite{Barausse:2011pu} it has been shown
that \ae-theory admits a one-parameter family of asymptotically 
flat, static, spherically symmetric solutions, 
which are regular everywhere apart from the central singularity 
and have a metric horizon (see also Ref.~\cite{Eling:2006ec} for earlier work). 
Since spherically symmetric vector fields are always hypersurface orthogonal, 
these are also solutions of the IR limit of HL gravity. 

These solutions present a metric horizon, which acts as a causal boundary for 
matter fields coupled minimally to the metric, and additional horizons, which act as 
causal boundaries for the gravity-sector modes. Thus, from an IR perspective 
these are indeed BHs. However, in HL 
gravity, dispersion relations cease to be linear at the scale $M_\star$, 
so short-wavelength perturbations travel at arbitrarily high speeds in 
the preferred frame and penetrate all horizons. One
expects the same to happen in any sensible UV completion of \ae-theory. 
Because $M_\star$ corresponds to a length scale
much smaller than the horizon of astrophysical BHs, 
one does not expect significant corrections to the solutions of Ref.~\cite{Barausse:2011pu}
near the horizons. Nevertheless, from a conceptual viewpoint, even though these solutions 
should be very close to those of the full UV-complete theory, 
the interpretation of the various horizons as true 
causal boundaries would be incorrect.
As a result, the very concept of a BH would not survive. 

A  striking feature of the solutions of Ref.~\cite{Barausse:2011pu} 
is the existence of a hypersurface that is orthogonal to the aether,
and which lies inside the various horizons and 
therefore cloaks the singularity. This hypersurface can 
be interpreted as a constant-time hypersurface as measured in the preferred frame defined by 
the aether (or as a constant $T$ hypersurface in HL gravity), and is also a constant $r$ hypersurface, 
where $r$ is the Schwarzschild coordinate.
This hypersurface acts as a {\em universal} 
horizon, as it constitutes a causal boundary for all modes, irrespectively of their propagation 
speed \cite{Barausse:2011pu}. The existence of this hypersurface has been shown also in 
Ref.~\cite{Blas:2011ni} in the decoupling limit. However, it has also been shown there that 
this universal horizon appears to be non-linearly unstable against non-spherically symmetric 
perturbations in the infrared limit of HL gravity (although not in \ae-theory). Thus, the question 
that we want to address is: do BHs really exist in HL gravity, if one abandons the 
idealized assumption of spherical symmetry? In this Letter we will focus
on slowly rotating BHs, which in GR can be obtained from the Schwarzschild solution
by expanding the field equations in the rotation parameter~\cite{hartle_thorne}.

We first show that the static, spherically symmetric, asymptotically flat 
solutions for \ae-theory exactly coincide with those of HL gravity, assuming that the aether $u^\mu$ in \ae-theory, and the scalar $T$ in HL gravity asymptote to their trivial configuration in Minkowski space, i.e. $u^\mu=\delta^\mu_t$ and $T=t$ in the preferred frame (see also Ref.~\cite{arXiv:1007.3503}).
Therefore, the solutions found in Ref.~\cite{Barausse:2011pu}
are the full set of static, spherical and asymptotically flat BHs of HL gravity.
This is not obvious because
the equivalence between \ae-theory and the infrared limit of HL gravity requires that the
 aether be hypersurface-orthogonal at the level of the action, and as a result
the two theories do not have the same field equations. In fact, variation of 
(\ref{actionae}) with respect to $u^\mu$ without assuming eq.~(\ref{hypers})  gives
a set of four equations $\AE_\mu=0$, whereas variation of (\ref{actionae}) with respect to 
$T$ yields
\begin{equation}\label{hleq}
\partial_\mu \left(\frac{1}{\sqrt{\nabla^\alpha T \nabla_\alpha T}} \sqrt{-g} \AE^\mu\right)=0\,.
\end{equation}
Solutions of $\AE^\mu=0$ satisfy also eq.~(\ref{hleq}), so hypersurface 
orthogonal solutions of \ae-theory will also be solutions of HL gravity. 
To show the converse, let us first note that once we impose spherical symmetry and staticity,
$\AE^\theta=\AE^\varphi=0$ identically, and $\AE^r=0$ implies $\AE^t=0$. Thus, it suffices to prove 
that eq.~\eqref{hleq}  yields $\AE^r=0$. 
Eq.~\eqref{hleq} now involves only $r$ and $\theta$ derivatives,
so integrating between $r=r_1$ and $r=r_2$ and using the divergence theorem yields
\begin{equation}
\int_{\theta=0}^{\theta=\pi}\frac{1}{\sqrt{\nabla^\alpha T \nabla_\alpha T}} \sqrt{-g} \AE^r d\theta\Bigg\vert^{r=r_2}_{r=r_1}=0
\end{equation}
Sending now $r_2\to\infty$,  from asymptotic flatness we get 
$\sqrt{\nabla^\alpha T \nabla_\alpha T}\sim 1$, 
$\sqrt{-g}\sim r^2\cos\theta$ and $\AE^r\sim \partial^2 u \sim 1/r^3$,
hence $\sqrt{-g} \AE^r/\sqrt{\nabla^\alpha T \nabla_\alpha T}\sim 1/r \to 0$. 
Therefore, at $r=r_1$ we must have $\AE^r=0$. 

We now turn our attention to rotating BHs. The most general slowly rotating, stationary, axisymmetric metric
can be written, in a suitable coordinate system, 
as~\cite{hartle_thorne}
\begin{eqnarray}\label{metric}
ds^2&=&f(r) dt^2 -\frac{B(r)^2}{f(r)}dr^2-r^2(d\theta^2+\sin^2\theta \,d\varphi^2)\nonumber\\
&&+\epsilon 
r^2 \sin^2\theta \,\Omega(r,\theta) dtd\varphi+{\cal O}(\epsilon^2)\,,
\end{eqnarray}
where $\epsilon$ is the book-keeping parameter  of the expansion. Since we are interested in slowly rotating BHs in HL gravity, $f(r)$ and $B(r)$ are given by the solutions of Ref.~\cite{Barausse:2011pu}.

As for $T$, given that it appears in the action only through $u_\mu$, 
 one cannot exclude a dependence on $t$ or $\varphi$, provided that 
 $\partial_t u_\mu=\partial_\varphi u_\mu=0$. In fact, given that $u^\mu$ 
 is  timelike a $t$-dependence is necessary. 
Hypersurface orthogonality
 for $u^\mu$ implies vanishing vorticity, and thus
yields $u_\varphi= \ell u_t $, where $\ell=$ constant can be interpreted as the angular momentum
of the aether per unit energy, as seen at infinity. Clearly, $\ell$ can be made to vanish if we make our coordinate
system corotate with the aether at infinity, i.e. if we perform the coordinate change $t'=t-\ell \varphi$, we have
$u_\varphi=0$. Provided that $\ell={\cal O}(\epsilon)$ (which follows from the slow-rotation assumption), it 
is easy to see that such a coordinate change leaves the ansatz (\ref{metric}) invariant modulo a redefinition of 
$\Omega(r,\theta)$. Without loss of generality we then set $u_\varphi=0$
and write
\begin{equation}\label{aether}\boldsymbol{u}= \frac{1+f A^2}{2 A}dt
+ \frac{B}{2A}\left(\frac{1}{f}-A^2\right) dr+{\cal O}(\epsilon)^2\,,
\end{equation}
 where we have also used $u_\theta={\cal O}(\epsilon^2)$
and $g^{\mu\nu}u_\mu u_\nu=1$.  $A$ is the aether component $u^v$ in ingoing Eddington-Finkelstein coordinates, used in Ref.~\cite{Barausse:2011pu}, and we have suppressed
the $r$-dependence of $A$, $B$ and $f$ to lighten the notation. Note, however, that in general $u^\varphi\neq0$. 

For $T$, the above 
translates into $T=t+\tau(r,\theta)$, if one also uses invariance under $T\to \tilde{T}(T)$. Assuming 
$u_\varphi\neq0$ would also imply a linear dependence of $T$ on
$\varphi$, and $T$ would not be single-valued. This would jeopardize the
equivalence between hypersurface orthogonal \ae-theory and HL gravity,
which requires that $T$ be identified with the time coordinate in the
preferred foliation.

The action (\ref{actionae}) remains unchanged under the set of field redefinitions $g'_{\alpha\beta}=g_{\alpha\beta}+(s^2-1)u_\alpha u_\beta$, $u'^\alpha=s^{-1} u^\alpha$, if the
$c_i$ are replaced with new couplings $\tilde{c}_i(c_i)$ \cite{Foster:2005ec}. $g'_{\alpha\beta}$ and $u'_\mu$ are still described by the ans\"atze \eqref{metric} and \eqref{aether} (after a suitable coordinate transformation).
For $s=s_0$, $s_0$ being the speed of the spin-0 mode, $g'_{\alpha\beta}$ becomes the effective metric on which spin-0 excitations propagate \cite{Eling:2006ec}. Hence, the horizon of $g'_{\alpha\beta}$  coincides with the causal boundary of the solutions of Ref.~\cite{Barausse:2011pu} for the spin-0 mode. We find it convenient to work with these redefined fields, as in Refs.~\cite{Eling:2006ec,Barausse:2011pu}. After the field redefinitions, one can still exploit hypersurface orthogonality for $u^\mu$ and set $\tilde{c}_4=0$.

\begin{figure}
\includegraphics[width=7cm,height=6.5cm]{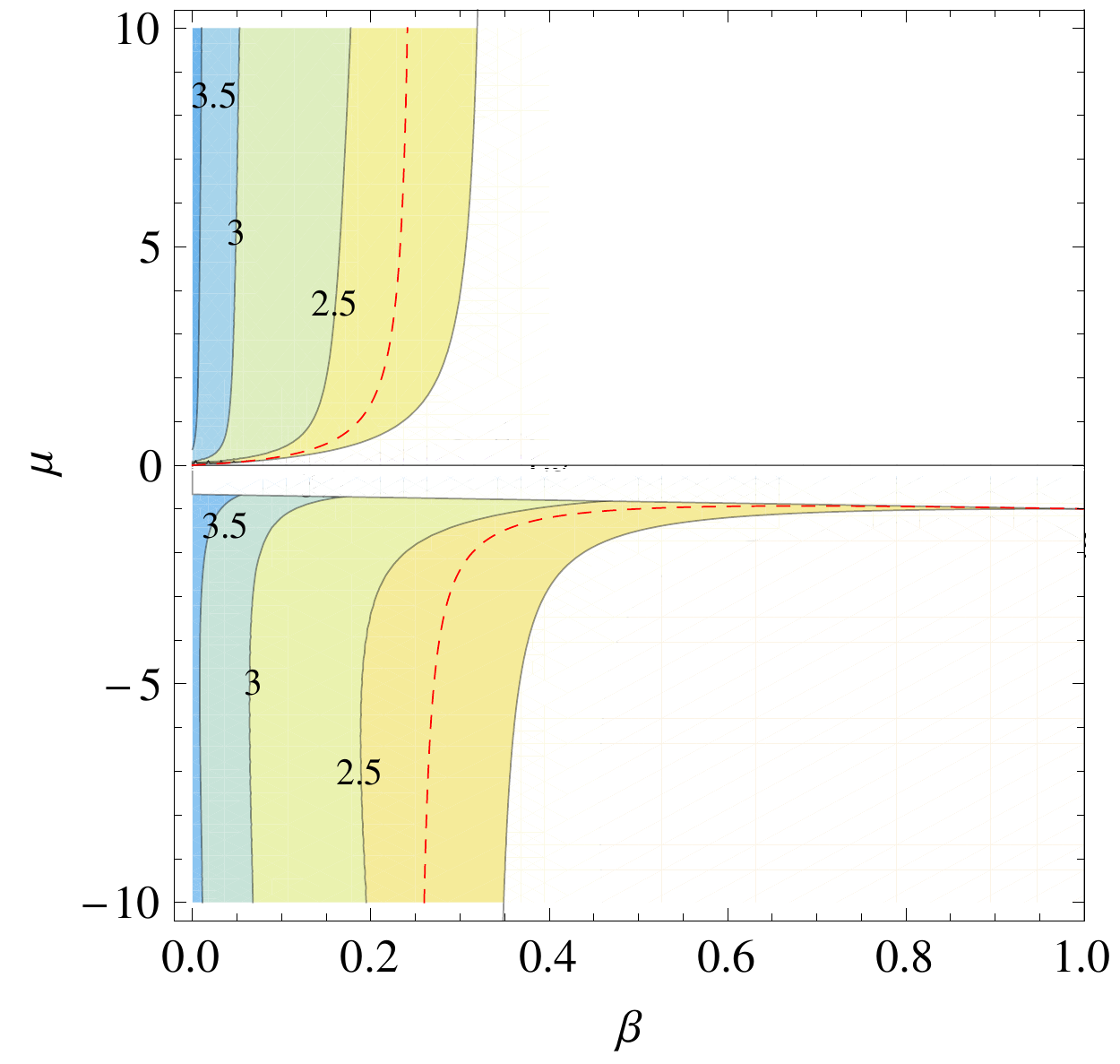}
\caption{\footnotesize Graphic representation of the viable part of the parameter space, where $\beta=(\xi-1)/\xi$ and $\mu=(\lambda-\xi)/\xi$. Contours represent the value of $S$. The dashed line represents $\tilde{c}_{13}=0$.\label{figure}}
\label{fig:EJ}
\end{figure}

As in Ref.~\cite{Barausse:2011pu},  we restrict ourselves to the part for the (physical) parameter space for which: 
(i) all propagating modes are stable and have positive energy, (ii) vacuum Cherenkov radiation by matter is avoided \cite{Elliott:2005va}, 
and (iii) there is complete agreement with GR at the first post-Newtonian order (vanishing preferred-frame parameters) \cite{arXiv:1007.3503}. 
The last condition requires that $\eta=2(\xi-1)$ and restricts the parameter space to two dimensions, see Fig.~\ref{figure}.
The corresponding static, spherically symmetric BHs, which act
as ``seeds'' for the slowly rotating BHs studied here, are presented in section VB of Ref.~\cite{Barausse:2011pu}.

At first order in $\epsilon$, there are three non-trivial independent HL field equations:
 \begin{align}
 \label{eq1}
&\frac{\tilde{c}_{13}}{8
    r^3 A^3 B f^2} 
 \Big\{f \Big[2 \partial_\theta\Omega (r,\theta) (A- r A')
+r A   \partial_r\partial_\theta\Omega (r,\theta)\Big]\nonumber\\&
 \quad -f^3 A^4 \Big[2  \left(r A'+A\right) 
 \partial_\theta\Omega (r,\theta)+r A \partial_r\partial_\theta\Omega (r,\theta)\Big]\nonumber\\&
  \quad -r A f' \partial_\theta\Omega (r,\theta) \Big(1+A^4 f^2\Big)\Big\}=0 \\
  & \frac{1}{r^2 f}\Big\{\frac{1}{2} 
\left[-\partial^2_\theta\Omega(r,\theta)-3 \cot\theta \partial_\theta\Omega(r,\theta)\right]
+
k_0 \Omega (r,\theta)\nonumber\\& \quad+k_1 \partial_r\Omega (r,\theta)
+k_2 \partial^2_r\Omega (r,\theta)\Big\}=0\label{eqtphi}\\
 & q_0 \Omega (r,\theta)+q_1 \partial_r\Omega (r,\theta)
+q_2 \partial^2_r\Omega (r,\theta)=0\label{eqrphi}
  \end{align}
where 
a prime denotes differentiation with respect to the argument. Also,
$k_i$, $q_i$, $i=0,1,2$ are functions of $\tilde{c}_1$ 
and $\tilde{c}_3$, as well as of $A$, $f$ and $B$ and their derivatives.

Now, $\Omega(r,\theta)=\Omega_0=$ constant must be a solution to these equations because it
is just the spherically symmetric static metric, transformed under the
coordinate change $\varphi'=\varphi+\Omega_0 t$. Thus, $k_0=q_0=0$ when one uses
the spherically symmetric static solution.

In the GR limit $\tilde{c}_1=\tilde{c}_3=0$, one has $q_1=q_2=0$, as well as
$k_1=r f \left(r B'-4 B\right)/(2 B^3)$
and $k_2=-r^2 f/(2 B^2)$. Using the Schwarzschild solution
$B=1$ and $f=1-2 M/r$, eq.~\eqref{eqtphi} then gives
\begin{multline}\label{GR}
-(r-2 M) \left[4 \partial_r\Omega(r,\theta)+r\,\partial^2_r\Omega(r,\theta)\right]\\
=
\partial^2_\theta\Omega(r,\theta)+3 \cot \theta \partial_\theta\Omega(r,\theta)\,.
\end{multline}
Solving by separation of variables and imposing regularity at the poles, one finds
the expected solutions $\Omega(r,\theta)=$ constant
and $\Omega(r,\theta)=\Omega_{\rm H} (2 M/r)^3$ ($\Omega_{\rm H}=$ constant being the horizon's angular
velocity), which is the slowly 
rotating limit of the Kerr solution.

In HL gravity instead there are extra equations, without any extra function to be determined. This is crucial as taking a linear combinations of 
eqs.~\eqref{eqtphi}--\eqref{eqrphi} can be used in order to eliminate $\partial^2_r\Omega (r,\theta)$ and obtain
\begin{align}\label{eq_combo}
&\frac{\tilde{c}_{13} \left(A^4 f^2-1\right)}{16 A^2 B^3}\Bigg\{\partial^2_\theta\Omega(r,\theta)
+3 \cot \theta \partial_\theta\Omega(r,\theta)\nonumber\\
&+2 \tilde{c}_1 r^2 (\tilde{c}_{13}-1) f
 \frac{\left(A^2 f-1\right) A'+A^3
   f'}{A B^2 \tilde{c}_{13} \left(A^2 f+1\right)}\,\partial_r\Omega(r,\theta)
\Bigg\}=0\,.
\end{align}
Solving naively by separation of variables and imposing
regularity at the poles, one finds $\Omega=$ constant (i.e.
the spherical solution in rotating coordinates) as the only solution. We
now show that this is indeed the only solution to eq.~\eqref{eq_combo}, even if one
does not assume $\Omega(r,\theta)=R(r)Q(\theta)$.
At the horizon $r_{\rm H}$, $f(r_{\rm H})=0$ but $f'(r_{\rm H})\neq 0$ and $A(r_{\rm H})\neq 0$~\cite{Barausse:2011pu},
 so eq.~(\ref{eq1}) yields
  $\partial_\theta\Omega (r_{\rm H},\theta)=0$, hence $\Omega (r_{\rm H},\theta)=\omega_0=$ constant. 
Because the spherical solution is regular at the horizon~\cite{Barausse:2011pu}, assuming analyticity,
we can write
\begin{gather}\label{f_series}
f(r)=\sum_{n=1}^{\infty}\frac{f^{(n)}(r_{\rm H})}{n!}(r-r_{\rm H})^n\,,\\
B(r)=B(r_{\rm H})+\sum_{n=1}^{\infty}\frac{B^{(n)}(r_{\rm H})}{n!}(r-r_{\rm H})^n\,,\\
A(r)=A(r_{\rm H})+\sum_{n=1}^{\infty}\frac{A^{(n)}(r_{\rm H})}{n!}(r-r_{\rm H})^n\,,
\end{gather}
and assuming that the slowly-rotating solution is also regular and analytic,
we can write
\begin{equation}\label{w_series}
\Omega(r,\theta)=\omega_0+\sum_{n=1}^{\infty}\omega_n(\cos\theta)(r-r_{\rm H})^n\,
\end{equation}
(Note that analyticity is also used to prove the uniqueness of the Kerr-Newman BH solution in GR~\cite{uniqueness}.)
At the lowest order in $r-r_{\rm H}$, eq.~(\ref{eq_combo}) gives 
\begin{equation}
\label{master_eq}
\frac{\tilde{c}_{13} [-4 \cos\theta \omega_1'(\cos\theta)+\sin^2\theta \omega_1''(\cos\theta)]}{2 \tilde{c}_1 (1-\tilde{c}_{13})} = 
S \omega_1(\cos\theta)\,,
\end{equation}
where 
\begin{equation}
\\
S\equiv\frac{ r_{\rm H}^2 f'(r_{\rm H}) \left[A(r_{\rm H})^3
   f'(r_{\rm H})-A'(r_{\rm H})\right]}{A(r_{\rm H}) B(r_{\rm H})^2}\,.
\end{equation}
Using the solutions of Ref.~\cite{Barausse:2011pu}, we have verified that $S\neq0$ 
in the viable regions of the $(\beta,\mu)$-plane (cf. Fig.~\ref{figure}).
In these regions, 
$\tilde{c}_1\neq0$, $\tilde{c}_{13}\neq1$, but there exist curves (shown in Fig.~\ref{figure}) 
on which $\tilde{c}_{13}=0$. If that is the case,
one immediately obtains $\omega_1(\cos\theta)=0$. In the rest of the viable regions, $\tilde{c}_{13}\neq0$
and the generic solution to Eq.~\eqref{master_eq} can be written in terms of hypergeometric functions,
\begin{align}
\omega_1(\cos\theta)=&\sigma_1 \, _2F_1\left(\frac{3-s}{4},\frac{3+s}{4} ;\frac{1}{2};\cos^2\theta\right)\\&+\sigma_2 \cos\theta \, _2F_1\left(\frac{5-s}{4},\frac{5+s}{4};\frac{3}{2};\cos^2\theta\right)\,,\nonumber
\end{align}
with $s\equiv \sqrt{9-8 \tilde{c}_1 (1-\tilde{c}_{13}) S/\tilde{c}_{13} }$. This general 
solution diverges at $\theta=0$ or $\theta=\pi$ unless $\sigma_1=\sigma_2=0$.
In order to prove that $\omega_n(\cos\theta)=0$ for any $n$, we can use the recursion theorem
and show that if $\omega_i(\cos\theta)=0$ for $i<n$, then $\omega_n(\cos\theta)=0$.
This follows from eq.~\eqref{eq_combo}, which at the lowest order in 
$r-r_{\rm H}$ (using $\omega_i(\cos\theta)=0$ for $i<n$) gives
\begin{equation}
\frac{\tilde{c}_{13} [-4 \cos\theta \omega_n'(\cos\theta)+\sin^2\theta \omega_n''(\cos\theta)]}{2 \tilde{c}_1 
(1-\tilde{c}_{13})} = 
n\,\omega_n(\cos\theta) S \,,
\end{equation}
from which we get that $\omega_n(\cos\theta)=0$, just like above.

In summary, we have shown that HL gravity does not admit stationary, axisymmetric, slowly rotating BHs. 
This is alarming because one expects a continuous 
limit from rotating to nonrotating BHs, as in GR.
Also, astrophysical BHs, for which there is nowadays robust evidence~\cite{narayan},
do have non-zero spins. 
More specifically, non-zero measurements
for the $g_{t\phi}$ component of the BH metric (``frame dragging''),
which our calculation predicts to be zero in HL gravity,
are provided by techniques such
as continuum fitting~\cite{continuum}
and analyses of the relativistic iron lines~\cite{iron}. 
Also, non-zero values of the spin (and thus of the frame dragging)
are naturally expected, at least for the ``massive'' BHs present in galactic centers, 
based on our current understanding of accretion and mergers during
 galaxy formation~\cite{spin_evolution}.
In particular, because the innermost stable circular orbit of
HL gravity non-rotating
BHs has a non-zero angular momentum~\cite{Barausse:2011pu}, thin-disk
accretion around these BHs
would naturally tend to spin them up.

A possible way out is that gravitational collapse never forms BHs in HL
gravity. While the
BHs of Ref.~\cite{Barausse:2011pu} have been found to form in a
perfectly spherical
collapse~\cite{garfinkle}, deviations from this idealized picture may
give rise to non-trivial
rotating configurations of matter and aether. However, for quasi-spherical initial conditions this seems unlikely, 
unless the higher order term of HL gravity (and the corresponding matter corrections) can somehow halt the collapse.
In any case, there is circumstantial evidence that astrophysical BH candidates possess an
event horizon
around them~\cite{narayan_horizon}, which would pose an additional problem
for such a scenario. Moreover, rotating ``BH mimickers'' that do not have an event
horizon  are
typically unstable classically due to
the so-called ergoregion instability~\cite{ergoregion}.

Note that in Refs.~\cite{Eling:2006ec,Barausse:2011pu} another 
set of static, spherically symmetric, asymptotically flat solutions 
with a metric horizon was found. These solutions were discarded because  
they exhibit a finite area singularity on the spin-0 horizon 
(the causal boundary of the spin-0 mode, which can be inside the metric horizon).  
It is indeed hard to imagine 
how such spacetimes may form from collapse, but one might 
 conceivably replace part of the 
interior with a configuration of matter and aether, so as to ``cover'' 
the singularity (cf. the discussion about the universal horizon 
instability in Ref.~\cite{Blas:2011ni}). These configurations
may have
a metric horizon surrounding the matter-aether configuration, and might replace the 
spherical BHs used as seeds for the slowly rotating 
solutions studied in this Letter.

Finally, 
 stationary, axisymmetric BHs might possibly exist, 
for which the aether does not share these symmetries (but its stress-energy tensor does). 
This may be similar to the ``stealth'' solutions of 2+1 gravity with a 
nonminimally coupled and self-interacting scalar field~\cite{stealth}.

{\em Acknowledgments}: We thank Ted Jacobson, Luis Lehner, Paolo Pani 
and Sergey Sibiryakov for many enlightening discussions and insightful comments of an earlier version of this manuscript.
EB acknowledges support from a CITA National Fellowship 
at the University of Guelph. TPS acknowledges partial financial
support provided under a Marie Curie Career Integration
Grant and the ``Young SISSA Scientists¢ Research
Project'' scheme 2011-2012. 

{\bf Note added}:  A subtlety in the dynamical equivalence between Ho\v rava--Lifshitz gravity and Einstein-aether theory has been missed and this has seriously affected the conclusions of this paper. Please see  arXiv:1212.1334 for a full discussion.

\end{document}